\documentclass[preprint,12pt]{elsarticle}

\usepackage{amssymb}

\usepackage{booktabs}
\usepackage{array}
\usepackage{longtable}
\usepackage{framed}
\usepackage[caption=false,font=footnotesize,farskip=0pt,captionskip=0pt]{subfig}

\begin{document}

\begin{frontmatter}



\title{An LLM-driven Scenario Generation Pipeline Using an Extended Scenic DSL for Autonomous Driving Safety Validation}


\author[inst1]{Fida Khandaker Safa\corref{cor2}}
\ead{fida.khandaker@students.mq.edu.au}
\affiliation[inst1]{organization={School of Computing, Macquarie University},
            city={Sydney},
            postcode={2109}, 
            state={NSW},
            country={Australia}}

\author[inst1]{Yupeng Jiang\corref{cor2}}
\ead{yupeng.jiang@hdr.mq.edu.au}

\author[inst1]{Xi Zheng\corref{cor1}}
\ead{james.zheng@mq.edu.au}

\cortext[cor1]{Corresponding author}
\cortext[cor2]{Equal contribution}

\begin{abstract}
Real-world crash reports, which combine textual summaries and sketches, are valuable for scenario-based testing of autonomous driving systems (ADS). However, current methods cannot effectively translate this multimodal data into precise, executable simulation scenarios, hindering the scalability of ADS safety validation. In this work, we propose a scalable and verifiable pipeline that uses a large language model (GPT-4o mini) and a probabilistic intermediate representation (an Extended Scenic domain-specific language) to automatically extract semantic scenario configurations from crash reports and generate corresponding simulation-ready scenarios. Unlike earlier approaches such as ScenicNL and LCTGen (which generate scenarios directly from text) or TARGET (which uses deterministic mappings from traffic rules), our method introduces an intermediate Scenic DSL layer to separate high-level semantic understanding from low-level scenario rendering, reducing errors and capturing real-world variability. We evaluated the pipeline on cases from the NHTSA CIREN database. The results show high accuracy in knowledge extraction: 100\% correctness for environmental and road network attributes, and 97\% and 98\% for oracle and actor trajectories, respectively, compared to human-derived ground truth. We executed the generated scenarios in the CARLA simulator using the Autoware driving stack, and they consistently triggered the intended traffic-rule violations (such as opposite-lane crossing and red-light running) across 2,000 scenario variations. These findings demonstrate that the proposed pipeline provides a legally grounded, scalable, and verifiable approach to ADS safety validation.

\end{abstract}




\end{frontmatter}


\section{Introduction} \label{sec:intro}
In safety‑critical domains such as autonomous driving, a single untested edge case can lead to catastrophic outcomes. Therefore, Autonomous Driving Systems (ADS) must be systematically exposed to the most hazardous and unusual scenarios before deployment \cite{Lou2021TestingOA, deng2022declarative, gladisch2019experience}. Scenario-based testing has become essential for validating ADS, as traditional brute‑force road testing is impractical \cite{zhong2021survey, hauer2019did}. Analyses show that an ADS would need to drive more than 11 billion miles to statistically demonstrate safety comparable to a human driver \cite{khastgir2021systems}. Instead, focusing on targeted driving scenarios can more efficiently reveal safety-critical weaknesses \cite{menzel2018scenarios, zhang2023realistic}. By simulating diverse and high-risk situations (e.g., near-collisions, sudden rule violations) that rarely occur in real-world data, scenario-based testing provides a practical path to ensuring robustness \cite{neurohr2020fundamental}.

However, capturing rare and complex real-world driving scenarios at scale remains challenging. Only a tiny fraction of recorded driving data involves the critical near-miss or accident situations most important for testing \cite{AutonomousVehicleAccidents}. Therefore, companies log vast miles and hire crash reconstruction experts to recreate rare events in simulation. For example, the National Highway Traffic Safety Administration’s Crash Injury Research Engineering Network (NHTSA CIREN) database contains over 2,500 detailed crash reports, each with narrative descriptions and scene diagrams \cite{NHTSA}. Transforming this wealth of multimodal crash data into executable test scenarios requires significant manual effort by experts. This manual approach lacks scalability and hinders the exploration of ``what‑if'' scenario variations. There is a clear need for automated methods to generate diverse, realistic crash scenarios that cover these corner cases at scale \cite{weber2022toward}.

Recent approaches have started leveraging large language models (LLMs) to generate driving scenarios from textual descriptions, but they face notable limitations. Directly prompting an LLM to produce scenario code (e.g., in simulation scripts like Scenic \cite{fremont2019scenic, fremont2023scenic} or OpenSCENARIO \cite{OpenSCENARIO}) often yields syntax errors or logically inconsistent scenarios \cite{Deng2023TARGETTR}. The complex grammar and constraints of scenario description languages make naive generation unstable, and LLMs may hallucinate details not present in the input. Methods such as LCTGen \cite{tan2023language} and ScenicNL \cite{elmaaroufi2024scenicnl} have proposed intermediate solutions. LCTGen uses an LLM (GPT-4) to translate a natural-language traffic description into a structured YAML scene specification, then generates traffic trajectories from it. This two-step process, along with techniques like chain-of-thought prompting, helps reduce hallucinations in LCTGen. Similarly, ScenicNL translates police crash reports into probabilistic Scenic programs by chaining multiple LLM prompts and integrating compiler feedback to ensure the code is syntactically correct. These efforts improve syntactic correctness but remain limited to textual inputs. More importantly, they fail to fully leverage the multimodal nature of real crash reports, such as the valuable spatial information in crash scene diagrams or sketches. Consequently, they cannot adequately capture the complex scene geometry and context present in real accidents.

In this paper, we propose an automated pipeline that addresses these challenges by using LLMs to extract structured scenario representations from multimodal crash reports. Our approach uses a custom LLM (GPT-4o mini \cite{GPT-4o-mini}) to parse each crash’s textual summary and accompanying diagram, producing a formal scenario specification in an Extended Scenic domain-specific language (DSL). A DSL describes scenarios; in this work, it builds on the Scenic probabilistic programming language for scenario modeling. We extend Scenic with additional constructs to form an intermediate probabilistic representation of the crash scenario. This Extended Scenic DSL serves as a middle layer between free-form inputs and the final simulation script. By decoupling high-level semantic extraction from low-level scenario generation, the pipeline confines the LLM’s output to a well-defined schema, which reduces hallucinations and syntax errors. The intermediate DSL captures uncertainty in the crash report (e.g., unknown exact speeds or positions) by representing scenario elements with probability distributions. This allows one crash description to yield a family of scenario variations rather than a single deterministic scene. The pipeline generates thousands of scenario variations from a single crash report, spanning environmental conditions, actor behaviors, and timing, while remaining consistent with the original incident. This probabilistic, multi-variation approach provides a scalable way to test ADS behavior over many randomized yet realistic reenactments of the same scenario.

We evaluated the proposed pipeline on real-world crash cases from the NHTSA CIREN database to assess both its extraction accuracy and its utility in simulation. For each case, we compared the Extended Scenic DSL scenario produced by our system against a manually curated ``golden'' scenario description (oracle). Results show that our LLM‑driven extraction achieves high semantic fidelity: it correctly captured all key environmental and road geometry attributes, and reproduced vehicle behaviors and trajectories with over 97\% accuracy against a human‑derived oracle. These accurate scenario representations were then used to automatically generate and execute tests in the CARLA simulator \cite{dosovitskiy2017carla} integrated with Autoware \cite{Autoware}. The executed scenarios successfully triggered the targeted safety-critical events in the simulated ADS. Across 2,000 random variations per scenario, the tests consistently induced the intended traffic‑rule violations described in the original crash reports, such as vehicles crossing into oncoming lanes or running red lights. This demonstrates that the pipeline can parse and formalize complex crash scenarios and convert them into effective, repeatable test cases that reliably reproduce real‑world failures in a safe simulation environment. The experimental findings suggest that our method offers a legally grounded and scalable approach to ADS safety validation, bridging the gap between written crash narratives and actionable simulation tests.

In summary, the key contributions of this work include:
\begin{itemize}
\item We propose a novel pipeline that employs an LLM to convert multimodal crash reports, including textual narratives and crash scene sketches, into precise scenario specifications. This framework automatically generates executable ADS test scenarios directly from real crash report data, integrating both text and diagrams.
\item We introduce an Extended Scenic DSL as an intermediate representation that separates semantic scenario understanding from concrete simulation scripts. This probabilistic DSL supports modular, compositional scenario descriptions, enabling syntax checking and reducing LLM hallucinations through structured output constraints. It also facilitates the generation of numerous scenario variations from a single report, improving test coverage for rare events.
\item We evaluate the pipeline on real crash cases, showing that it achieves near‑oracle accuracy in extracting scenario semantics and produces scenarios that reliably recreate targeted traffic rule violations in CARLA with Autoware. Consistently triggering known failure modes across thousands of runs demonstrates the effectiveness of the generated scenarios in uncovering safety‑critical issues.
\end{itemize}

The remainder of this paper is structured as follows. Section~\ref{sec:method} details the design of our pipeline, including the GPT-4o mini prompting strategy and the Extended Scenic DSL syntax and semantics. Section~\ref{sec:experiment} describes the experimental setup, the crash report dataset, and the validation methodology using a golden oracle and simulator-in-the-loop testing. Section~\ref{sec:results} presents the evaluation results and analysis of the system’s performance. Section~\ref{sec:related_work} reviews related work on scenario-based testing and prior approaches to scenario generation with LLMs and other AI techniques. Section~\ref{sec:discussion} discusses the implications of our findings and limitations of the current approach. Finally, Section~\ref{sec:Conclusion} concludes the paper.

\section{Methodology} \label{sec:method}
\subsection{Overview}
\begin{figure*}[!ht]
	\setlength{\belowcaptionskip}{-0.3cm}
	\centering
	\includegraphics[width=0.9\textwidth]{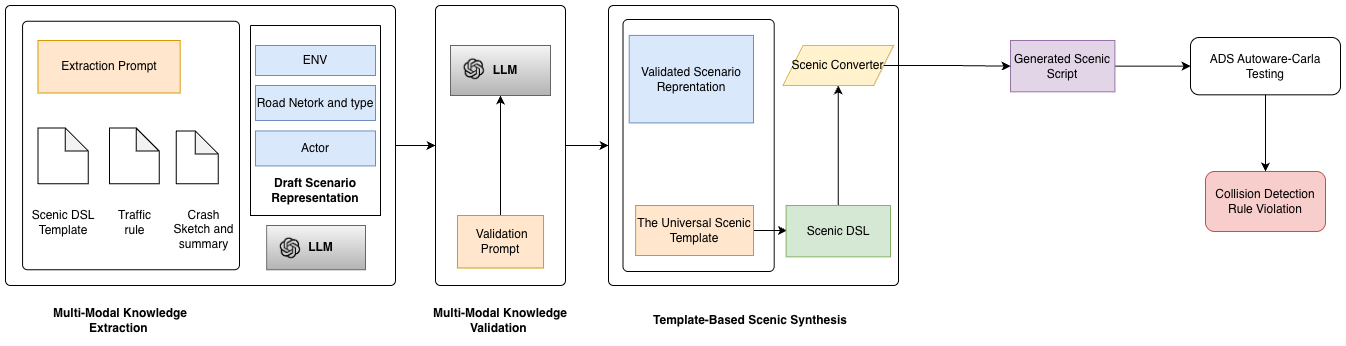}
	\caption{An overview of the full pipeline architecture.}
	\label{pipeline}
\end{figure*}

We extend the TARGET framework \cite{Deng2023TARGETTR} to scale autonomous driving system validation from formal traffic rules to real-world crash reports, with the goal of testing many plausible variants of a reported crash rather than a single deterministic reenactment. Figure~\ref{pipeline} shows the full pipeline architecture. In TARGET, the system extracts a test scenario from a traffic rule, encodes it in a deterministic DSL, and converts the result into a CARLA-oriented execution format. This rule‑focused design supports verification, but it depends on fixed coordinates and predefined map positions, and therefore cannot represent the uncertainty and incomplete details that often appear in crash reports. Our extension introduces an Extended Scenic DSL and an LLM-based knowledge extraction and validation pipeline that ingests multimodal crash reports, including textual narratives and crash sketches. The pipeline first performs high-level semantic extraction and validation, then produces a structured intermediate representation in the Extended Scenic DSL. Unlike the TARGET DSL, which is mainly designed for deterministic scenario reproduction, the Extended Scenic DSL acts as a probabilistic intermediate layer that represents scenario elements as distributions, which allows uncertainty modeling and controlled scenario variation. To turn this validated representation into executable tests, we add a template-based Scenic synthesis stage. This stage keeps semantic parsing separate from low-level rendering and converts the Extended Scenic DSL into Scenic programs through a topology-aware template engine. A universal Scenic template is used as the main blueprint, and it is instantiated according to road topology and scenario semantics, such as Straight, Curve, Intersection, and T-Intersection cases. A Scenic converter then maps the validated DSL fields into template parameters, normalizes tokens, and applies default values when crash reports omit details, which improves generation stability. The synthesis stage also supports parameter variation in key factors such as vehicle speed, position, and timing, so the framework can search across crash-consistent scenario variants and identify conditions that trigger unsafe ADS behavior. By combining probabilistic representation with template-based Scenic generation, the extended framework improves scalability, preserves crash semantics, and expands test coverage beyond single scenario replay.


\subsection{Extended Scenic DSL}
To bridge from abstract rule descriptions to concrete simulations, we introduce an Extended Scenic-oriented DSL for scenario generation. The Extended Scenic DSL is a structured intermediate language that encodes crash and traffic-rule scenario knowledge as semantic tokens, enabling scalable scenario synthesis for ADS testing without requiring low-level simulator parameters. The DSL is YAML‑based and defined by a context‑free grammar. A scenario consists of environment, road network, actors, and oracle.

\begin{itemize}
\item \textbf{\textit{Environment}} captures atmospheric and temporal context using categorical values such as sunny, cloudy, overcast, rainy, snowy, foggy, windy, or not mentioned, along with daytime, nighttime, or not mentioned. These values are later mapped to fixed weather and time presets.
\item \textbf{\textit{Road network}} describes topology, not absolute coordinates, using primitives such as straight, intersection, T‑intersection, and curve, along with structural fields that include number of ways and lanes (positive integers), road markers (e.g., solid or broken lines), and traffic signs (e.g., stop sign, speed limit sign, or not mentioned).
\item \textbf{\textit{Actors}} are defined as an ego vehicle with optional NPC actors, each including a type (car or truck), behavior (e.g., go forward, turn left, static), and position specified through a position reference, spatial relation (front, behind, left, right), and heading relation (same\_direction, opposite\_direction, from\_left, from\_right). This extends the original TARGET DSL by replacing explicit waypoint lists with behavior intent and relative positioning.
\item \textbf{\textit{Oracle}} is expressed as one or more \texttt{CVC\_$<$code$>$: $<$violation\_type$>$} entries, each accompanied by a quoted natural‑language description, with \texttt{CVC} denoting the California Vehicle Code \cite{VEH}.
\end{itemize}

\begin{figure*}[!ht]
	\setlength{\belowcaptionskip}{-0.3cm}
	\centering
	\includegraphics[width=0.8\textwidth]{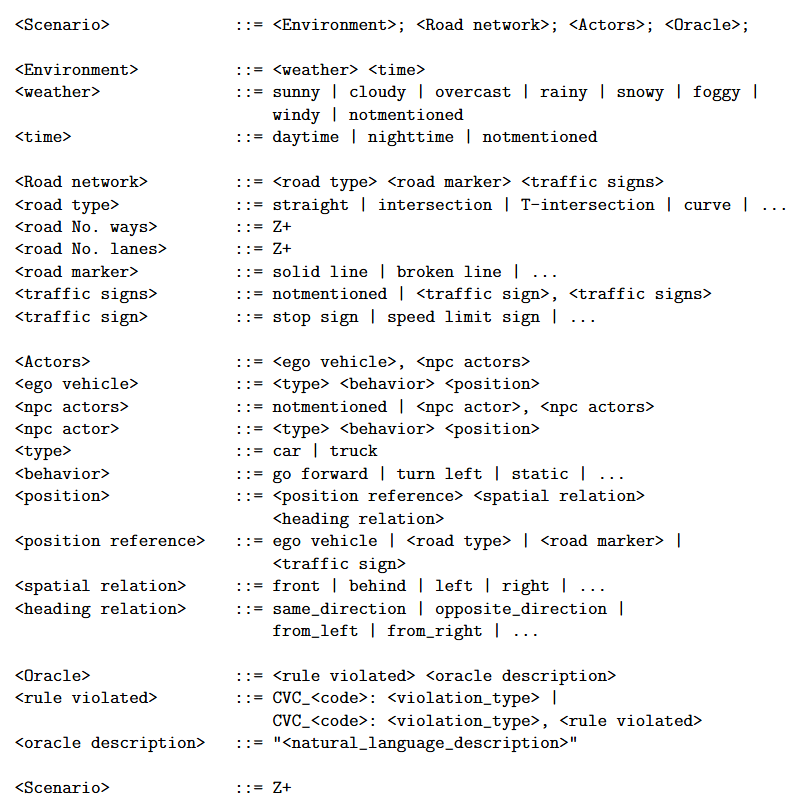}
	\caption{Definition of Extended Scenic DSL.}
	\label{DSL}
\end{figure*}

The Extended Scenic DSL supports probabilistic scenario synthesis by treating selected elements as random variables with distributions when compiled into Scenic, and using templates with placeholders to defer concrete parameter choices. In contrast, TARGET uses its DSL to reproduce traffic‑rule scenarios in a primarily deterministic manner. It maps abstract elements to concrete simulator parameters through lookup and map‑based search, rather than generating probabilistic Scenic programs. This probabilistic compilation enables systematic parameter exploration for ADS testing through sampling or search. These template‑based renderings transform the DSL into runnable Scenic programs and execute them in the CARLA simulator with Autoware controlling the ego vehicle. During each run, a real‑time scenario monitor validates behavior against 11 California traffic regulations using quantitative checks such as stopping behavior, speed compliance, lane keeping, and headway distance. It records violations when conditions are not met and classifies outcomes as rule violations or collisions, keeping results both executable and diagnostically useful. Figure~\ref{DSL} shows the definition of Extended Scenic DSL.

\subsection{LLM-based Multimodal Knowledge Parsing}
\subsubsection{Knowledge Extraction}
\begin{figure*}[!ht]
	\setlength{\belowcaptionskip}{-0.3cm}
	\centering
	\includegraphics[width=0.9\textwidth]{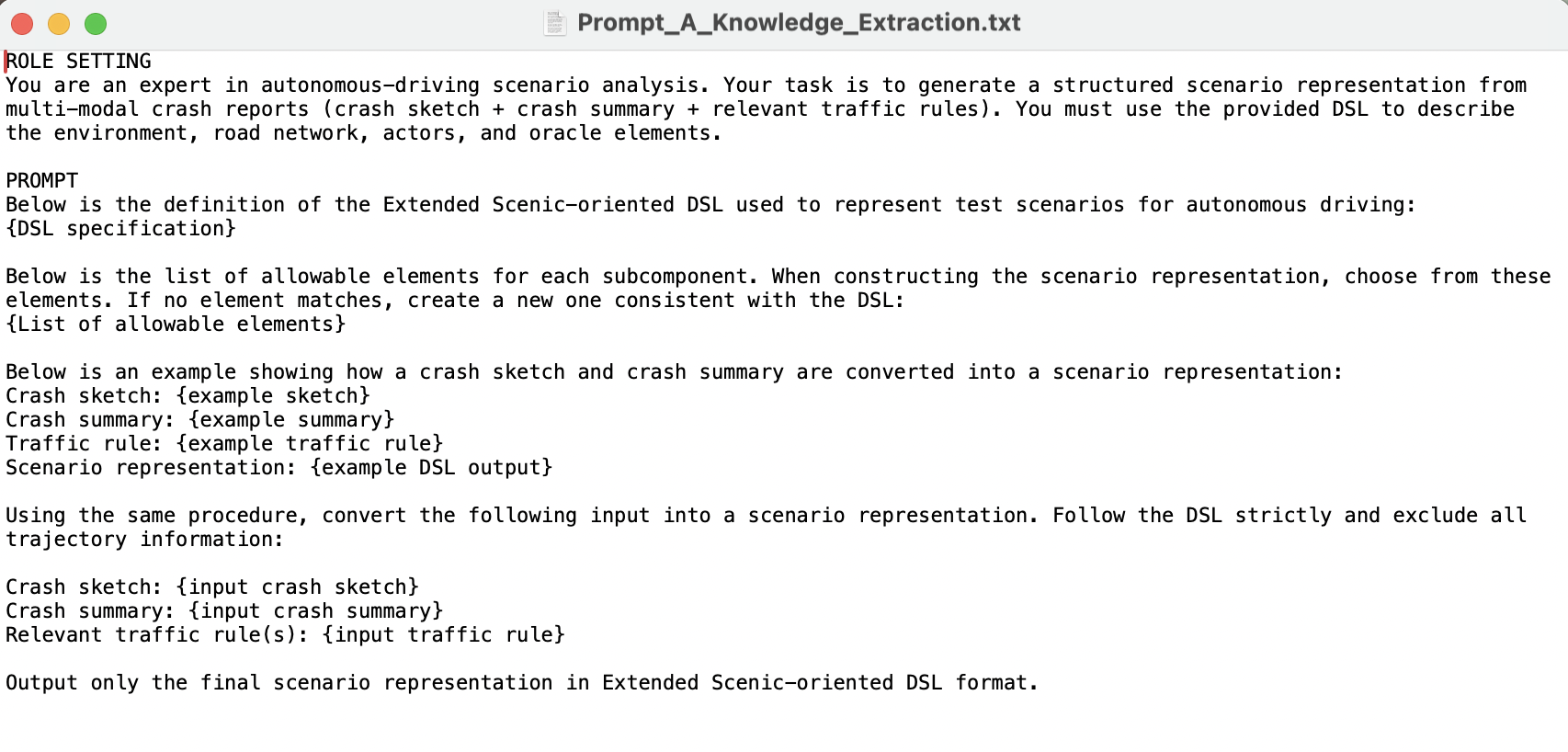}
	\caption{Extraction Prompt.}
	\label{Extraction}
\end{figure*}

Our pipeline parses multimodal crash reports into the Extended Scenic DSL by using an LLM as a structured extractor. The extraction stage uses GPT‑4o mini to convert unstructured inputs into a draft DSL representation. It jointly processes the crash sketch, crash summary, and relevant traffic regulations to infer environment, road network, and actor attributes. To keep the output aligned with the DSL ontology, we adopt few-shot prompting with exemplar pairs of crash descriptions and their correct DSL outputs (illustrated in Figure~\ref{Extraction}), which act as anchors that reduce ambiguity during generation. Instead of asking the model to produce a full YAML instance in one step, we guide it through a sequence of constrained decisions. This extraction prompt specifies the DSL schema, enumerates allowed element types, and requires the model to ground each extracted attribute in the provided evidence sources. The result is a draft scenario representation that remains interpretable and consistent with the Extended Scenic DSL specification.

\subsubsection{Knowledge Validation}
\begin{figure*}[!ht]
	\setlength{\belowcaptionskip}{-0.3cm}
	\centering
	\includegraphics[width=0.9\textwidth]{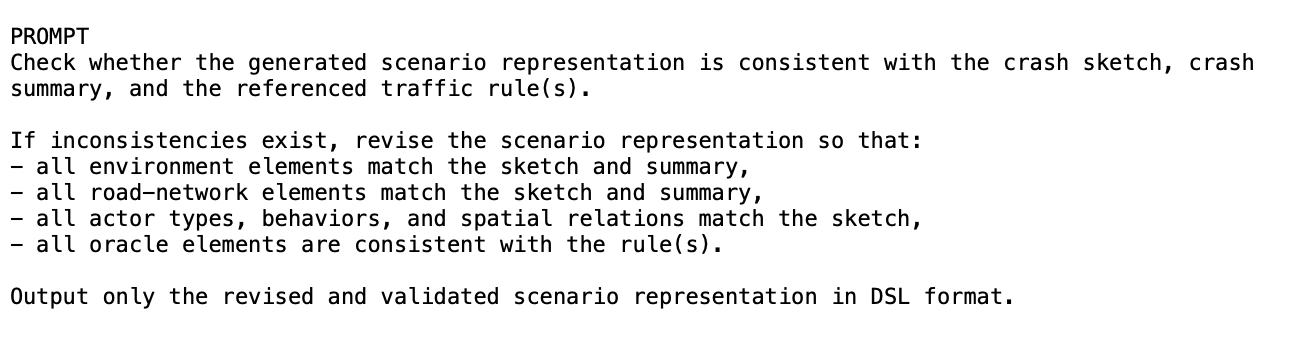}
	\caption{Validation Prompt.}
	\label{Validation}
\end{figure*}

After the draft representation is produced, the pipeline performs an integrated validation stage rather than a separate post-processing step. The model performs a semantic self‑check to verify that each DSL field is logically consistent and supported by evidence from the crash sketch, crash summary, and applicable traffic rules. For each field, the validation prompt (Figure~\ref{Validation}) checks whether the assigned value is grounded in textual or visual evidence. When inconsistencies occur such as an incorrect lane count, an unmentioned weather condition, or an actor behavior that conflicts with the sketch, the model revises the DSL accordingly. This step yields a validated representation in which each component is directly supported by the original crash documentation, which helps limit unsupported inference and reduces hallucination.

\subsection{Template-based Scenic Synthesis}
Template-based Scenic synthesis is the stage that converts the validated DSL representation into an executable Scenic program for simulation. In this stage, the method employs a topology‑aware Scenic template engine to translate a functional scenario description (such as a straight road or intersection) into the concrete simulation requirements of CARLA, including map assets and coordinate constraints. Unlike the original TARGET framework, which depends on runtime coordinate search, this design uses a deterministic template system. The template system organizes scenarios into four road topologies, namely \textit{Straight}, \textit{Curve}, \textit{Intersection}, and \textit{T-Intersection}, so that geometric validity is enforced during generation.

The core of this stage is a universal Jinja2-based \cite{ronacher2008jinja2} template that acts as a master Scenic blueprint. The template adapts its structure from the \texttt{Road\_Network} fields in the DSL and performs three translation functions.

\begin{itemize}
\item It maps environment and map information into simulator-ready values, including time conversion, weather preset mapping, and map selection for different road types. For example:
    \begin{itemize}
    \item \textbf{Time:} Abstract values map to specific simulation hours: Daytime $\rightarrow$ 12:00, Nighttime $\rightarrow$ 22:00.
    \item \textbf{Weather:} DSL descriptors are converted into CARLA‑compatible presets. For example, Clear weather and Night time map deterministically to the \texttt{ClearNight} preset, while Cloudy and Day become \texttt{CloudyNoon}.
    \item \textbf{Map Selection:} The template automatically selects the appropriate map asset based on the road configuration to ensure the required geometry exists: Straight (2-Lane) $\rightarrow$ \texttt{Town02} (rural environment), Straight (4-Lane) $\rightarrow$ \texttt{Town04} (highway environment), Curve (2-Lane) $\rightarrow$ \texttt{Town02} (rural environment), Intersections/T-Intersections $\rightarrow$ \texttt{Town05} (urban grid).
    \end{itemize}
\item It injects topology-specific logic blocks.
    \begin{itemize}
    \item \textbf{Straight Road:} handles \textit{head-on} and \textit{car-following} configurations based on heading relation.
    \item \textbf{Curve:} selects road segments with sufficient heading change.
    \item \textbf{Intersection/T-Intersection:} instantiate valid 4-way or 3-way junctions with conflict paths for the ego and non-player vehicles.
    \end{itemize}
\item It adds parameter variability through \texttt{VerifaiRange} constraints, so the generated scenario can sample from distributions instead of using fixed values (e.g., a base speed of 10 m/s becomes \texttt{VerifaiRange}(8, 12)). It also supports robust testing as speed and initial distance can vary within controlled ranges while the scenario semantics remain unchanged (e.g., \texttt{EGO\_INIT\_DIST} with \texttt{NPC\_INIT\_DIST} set to [15, 20] meters).
\end{itemize}

A Python-based Scenic converter is used as middleware between the YAML DSL and the Scenic template. The converter normalizes diverse textual inputs into canonical tokens, extracts heading relations to select the correct lane placement logic, and applies fallback defaults when crash reports omit details.

\begin{itemize}
\item \textbf{Data Normalization:} standardizes diverse text inputs into canonical tokens (e.g., ``Day'', ``Time'', ``12 pm'' $\rightarrow$ \texttt{time\_hour=12}).
\item \textbf{Heading Relation Extraction:} parses the heading relation field (e.g., \texttt{opposite\_direction} vs. \texttt{same\_direction}) to determine the adversary’s lane placement logic.
\item \textbf{Hard-coded Defaults (Missing Data Handling):} injects hard-coded defaults to ensure that every scenario remains executable:
    \begin{itemize}
    \item \textbf{Default Speed:} 10 m/s (approx 36 km/h) if unspecified.
    \item \textbf{Default Ego:} \texttt{vehicle.lincoln.mkz\_2017}.
    \item \textbf{Non-player Vehicle Models:} A vehicle is selected at random from a pool of common models such as \texttt{vehicle.nissan.patrol}, \texttt{vehicle.tesla.model3}, \texttt{vehicle.dodge.charger\_2020}, etc. If the actor type is specified as ``truck'', the default model is set to \texttt{vehicle.carlamotors.european\_hgv}.
    \item \textbf{Default Relation:} \texttt{opposite direction} (highest risk configuration).
    \end{itemize}
\end{itemize}

This design improves stability and reduces failure in script generation. As a result, the synthesized Scenic scripts are syntactically correct, topologically consistent with the original crash description, and suitable for repeated ADS testing under controlled variability.

\section{Experiment}\label{sec:experiment}
\subsection{Research Questions}
This study evaluates whether an LLM can generate precise scenario representations from multimodal crash reports and support closed-loop ADS testing. The main research question asks how to use an LLM to generate precise scenario representations from crash summaries and sketches, and it is refined into three sub-questions:
\begin{itemize}
\item \textbf{RQ1 (Representation Accuracy)} examines how accurately the Extended Scenic DSL captures semantic and spatial information compared with a human-derived golden oracle;
\item \textbf{RQ2 (Scenario Fidelity)} evaluates whether the generated Scenic scenarios reconstruct the dynamics and geometric structure of the original crash sketches;
\item \textbf{RQ3 (Testing Utility)} investigates what failures or traffic-rule violations emerge when the generated scenarios are executed in an Autoware-based autonomous driving system.
\end{itemize}

\subsection{Crash Cases and Traffic Rule Oracles}
To avoid unrealistic situations and ground scenarios in real crashes, we used the NHTSA CIREN case viewer database \cite{NHTSA, NHTSA_injury, NHTSA_report}, which contains 2,538 cases, and sampled 100 cases for the experiment due to time constraints. To connect crash reconstructions with lawful driving expectations, we integrated a curated set of California traffic regulations and aligned them with the scenario logic. The rules were derived from the California Driver’s Handbook \cite{handbook} and cross-referenced with the California Vehicle Code (CVC) \cite{VEH} to maintain legal consistency and simulation fidelity. The integrated rule set covers core domains that are common in safety-critical crashes, including right-of-way and intersection priority (CVC \S\S~21800--21804), signal and sign compliance (e.g., red lights CVC \S~21453 and stop signs CVC \S~22450), speed regulations (CVC \S\S~22349--22350), overtaking restrictions (CVC \S\S~21460--21461), and lane maneuvers (CVC \S\S~22107--22108). These regulations also serve as rule oracles during evaluation, where each scenario is assessed against explicit statutory conditions such as yielding at intersections and stopping at a red light.

\subsection{LLM Selection and ADS Testbed}
GPT‑4o mini \cite{GPT-4o-mini} was selected as the LLM for its support of multimodal inputs at a practical cost. It also demonstrates stronger multimodal reasoning performance compared to Gemini Flash and Claude Haiku. For ADS testing, we executed the generated Scenic scenarios in a closed-loop setup with Autoware as the ADS under test and CARLA as the simulator. We used two integration methods to connect Autoware and CARLA: the CARLA–Autoware ROS bridge \cite{CarlaAutowareBridge} and the Zenoh bridge \cite{Zenoh} provided by the Autoware Zenoh integration workflow. In the ROS bridge setup, CARLA 0.9.15 ran in a GPU‑enabled Docker container with host networking and a configured RPC port. A Carla Autoware Bridge container managed the connection and simulation parameters. Autoware Universe ran from the recommended tagged Docker image, using the carla\_t2 vehicle model, sensor kit, and lanelet2‑format maps supplied by the bridge project. The system followed the startup order CARLA, then the bridge, then Autoware to ensure stable communication. For the Zenoh bridge configuration, we installed Docker and Rocker to provide a reproducible container runtime for both the Zenoh bridge and Autoware. We cloned the \texttt{autoware\_carla\_launch} repository on the Humble branch and used it to build both the Zenoh CARLA bridge and the Autoware containers. The bridge-side container was built and launched first, using scripts that prepared and compiled the Zenoh CARLA bridge and then connected it to a running CARLA instance (version 0.9.13). Autoware containers were built and started on a second host, or on the same host with isolation, using dedicated scripts that configured the environment and compiled the required Autoware packages. In the demo setup, CARLA and the Zenoh CARLA bridge ran on one host, while two Autoware containers and Zenoh bridge DDS instances ran on another host to represent two independent vehicles. Each Autoware container used ROS localhost only to constrain DDS communication within the container, and vehicle-specific launch scripts were called with unique vehicle identifiers and the IP address of the host running the bridge. Using both bridge configurations enabled a broader evaluation of Autoware on Scenic‑generated scenarios and strengthened the assessment of scenario fidelity and ADS behavior. The implementation\footnote{https://github.com/FidaKhandaker/ThesisB\_code} transforms the DSL into runnable Scenic programs and executes them in the CARLA simulator with Autoware controlling the ego vehicle.

\section{Results}\label{sec:results}
This section reports the results for the three research questions. We first evaluate the accuracy of the LLM based knowledge representation in the Extended Scenic DSL (RQ1). We then assess whether the generated scenarios preserve the semantics of the original crash reports (RQ2). Finally, we examine whether the generated scenarios are useful for ADS testing in Autoware and whether they can reliably trigger meaningful traffic rule violations and safety critical failures (RQ3).

\subsection{RQ1: Representation Accuracy}
\begin{figure}[!ht]
    \centering
    \subfloat[LLM extracted Extended Scenic DSL for scenario 128697.]{\includegraphics[width=6cm]{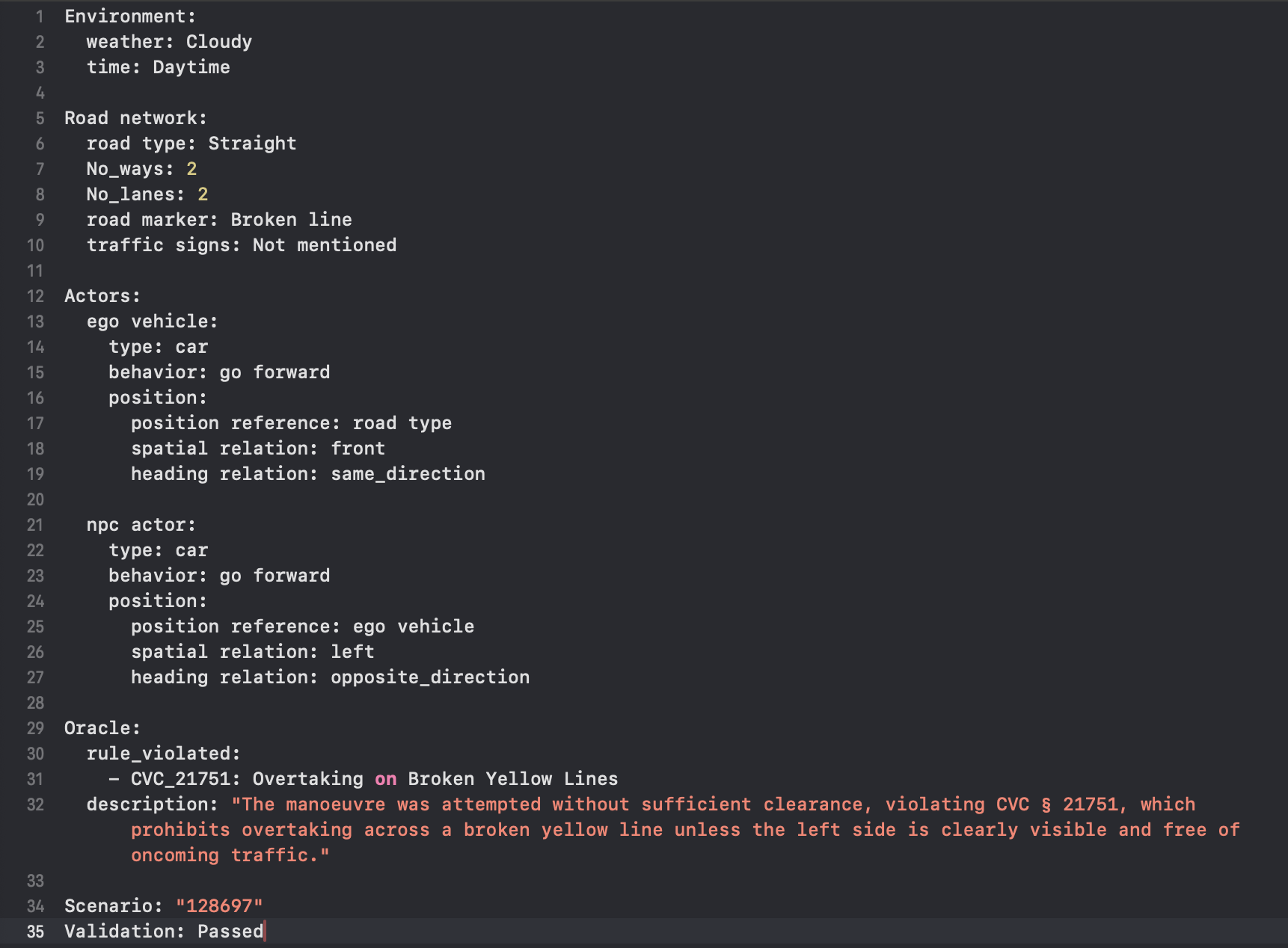}
    \label{fig:LLM_DSL}}
    \hspace{1cm}
    \subfloat[Manually written golden oracle for scenario 128697.]{\includegraphics[width=6cm]{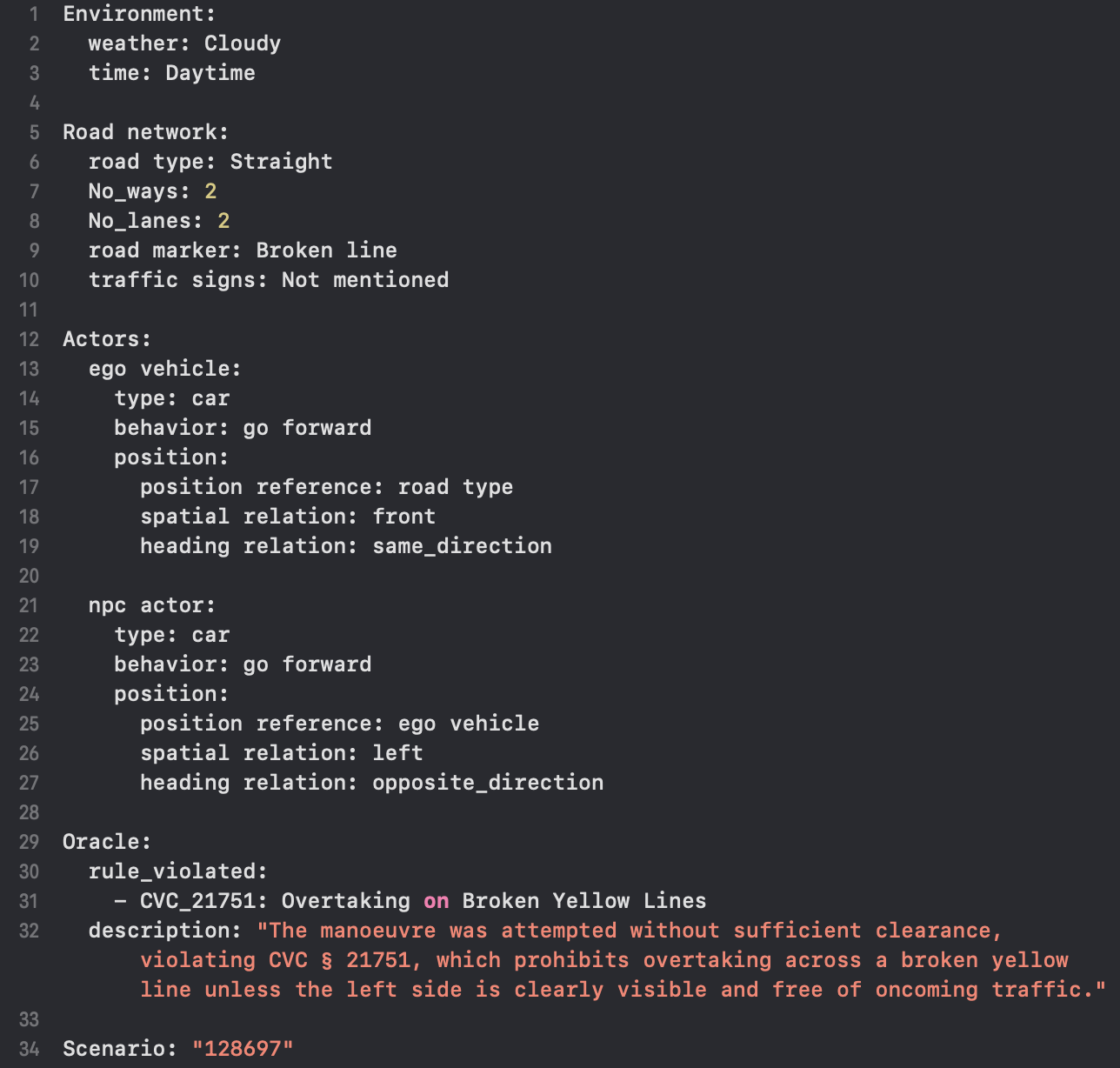}
    \label{fig:golden_oracle}}
    \captionsetup{font=small}
    \caption{Illustrative side-by-side comparison for the same crash scenario.}
    \label{fig:side-by-side}
\end{figure}

To answer RQ1, we compared the LLM extracted Extended Scenic DSL against a manually written golden oracle for 50 cases. Figure~\ref{fig:side-by-side} presents a side‑by‑side comparison of the same crash scenario between the LLM extracted Extended Scenic DSL and a manually written golden oracle. The results show strong overall performance, with an overall accuracy of 99\% across the DSL subcomponents, which indicates that the proposed extraction and validation pipeline can produce reliable structured representations from crash reports.

For this evaluation, inter-rater agreement is a key validity requirement because the golden oracle is human derived. A human oracle serves as a reliable reference only when annotators demonstrate strong agreement on the same cases. In the TARGET study, the authors explicitly used a weighted linear Fleiss Kappa \cite{fleiss1971measuring} and interpreted a value of 0.68 as substantial agreement, based on the standard interpretation range of 0.61 to 0.80 \cite{landis1977measurement}. We follow the same principle here, namely that kappa based agreement should be reported to justify the use of the human derived oracle as the reference for RQ1.

\begin{figure*}[!ht]
	\setlength{\belowcaptionskip}{-0.3cm}
	\centering
	\includegraphics[width=0.9\textwidth]{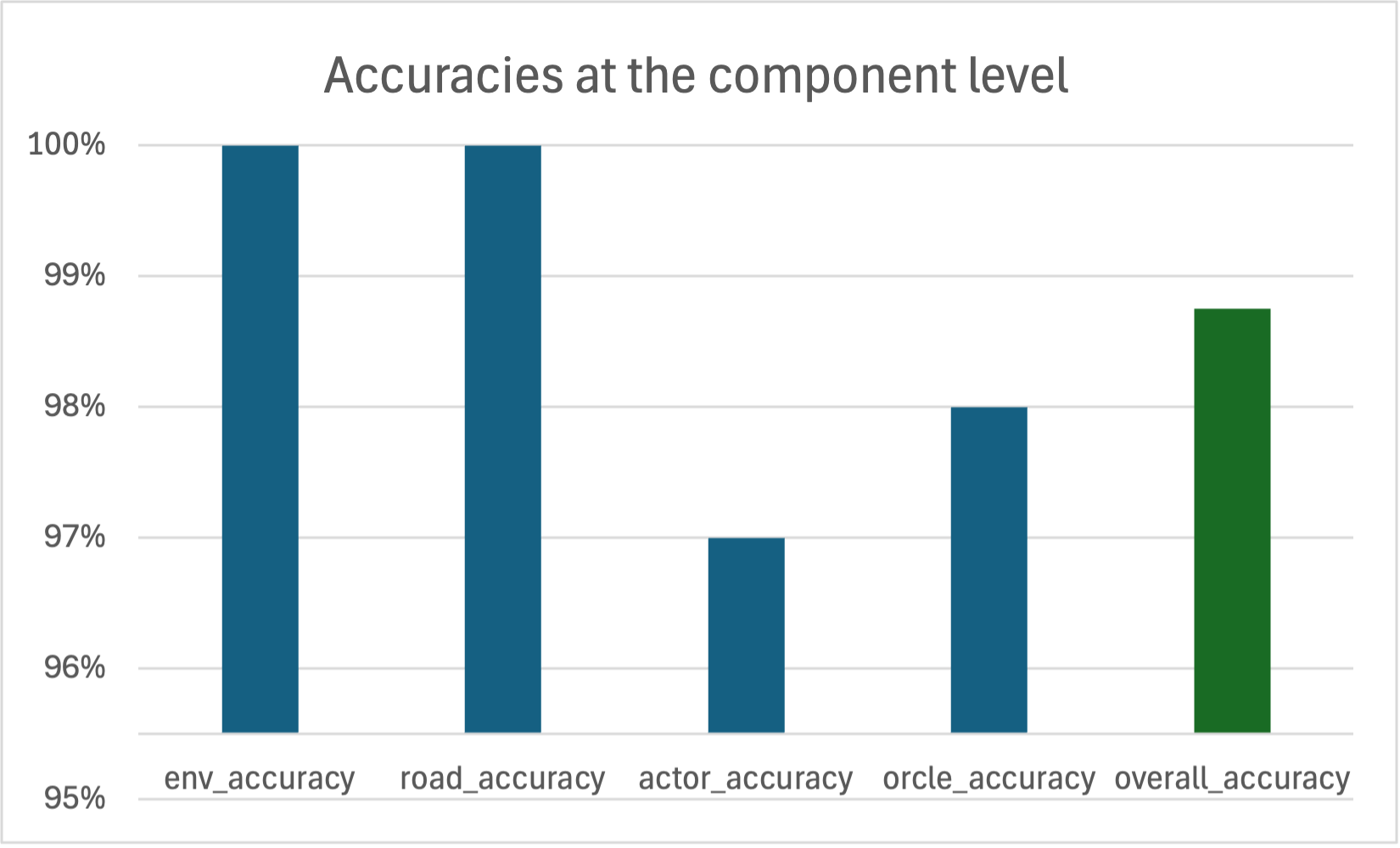}
	\caption{LLM extracted Extended Scenic DSL representation component accuracy scores against golden oracle.}
	\label{component}
\end{figure*}

The component level results also show a clear pattern (Figure~\ref{component}). Environment and Road Network reached 100\% accuracy, which suggests that the model is highly reliable on static and categorical attributes. The Oracle component reached 98\%, and the Actor component reached 97\%. The remaining errors stem mainly from two sources. First, traffic rules often include qualitative legal language, which is harder to map into precise formal logic. Second, police sketches sometimes contain ambiguous arrows or visually noisy trajectories, which makes actor behavior extraction harder than extracting static road context. These findings support the robustness of the representation while also showing where future improvement is most needed.

\subsection{RQ2: Scenario Fidelity}
\begin{figure}[!ht]
    \centering
    \subfloat[Scenario 128697 simulated scene in CARLA simulator.]{\includegraphics[width=6cm]{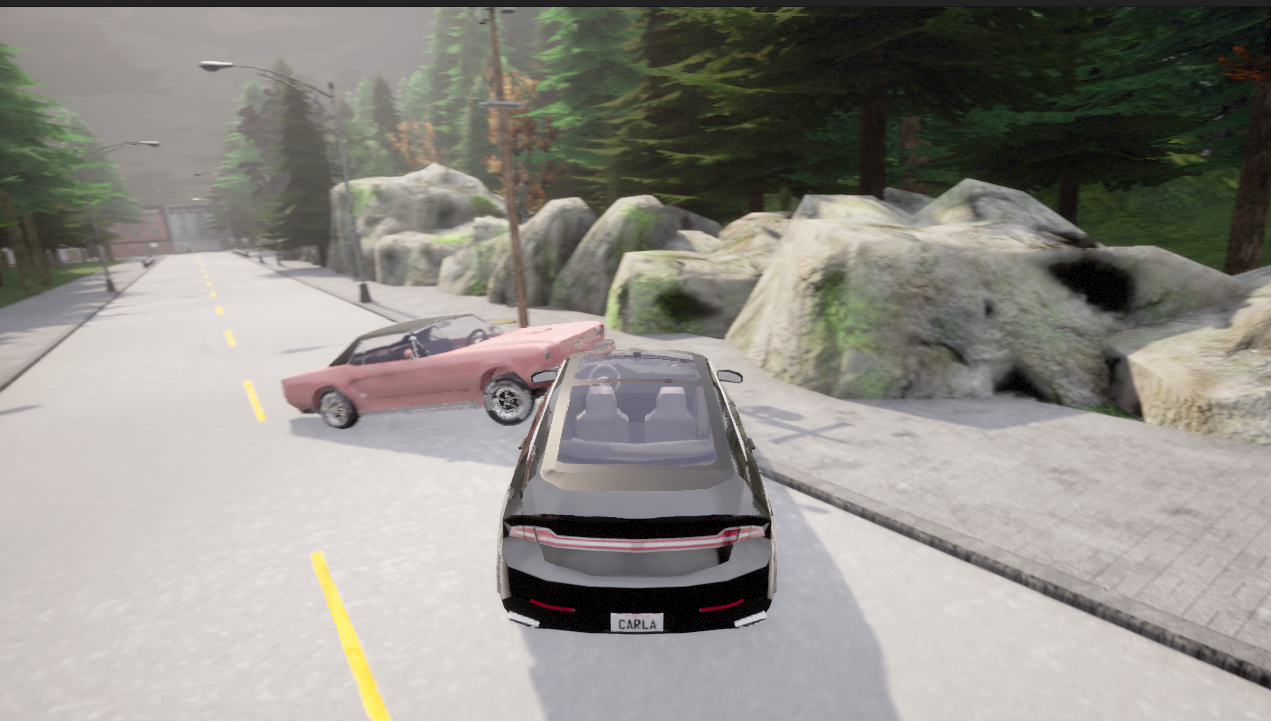}
    \label{fig:128697_CARLA}}
    \hspace{1cm}
    \subfloat[Original sketch of scenario 128697.]{\includegraphics[width=6cm]{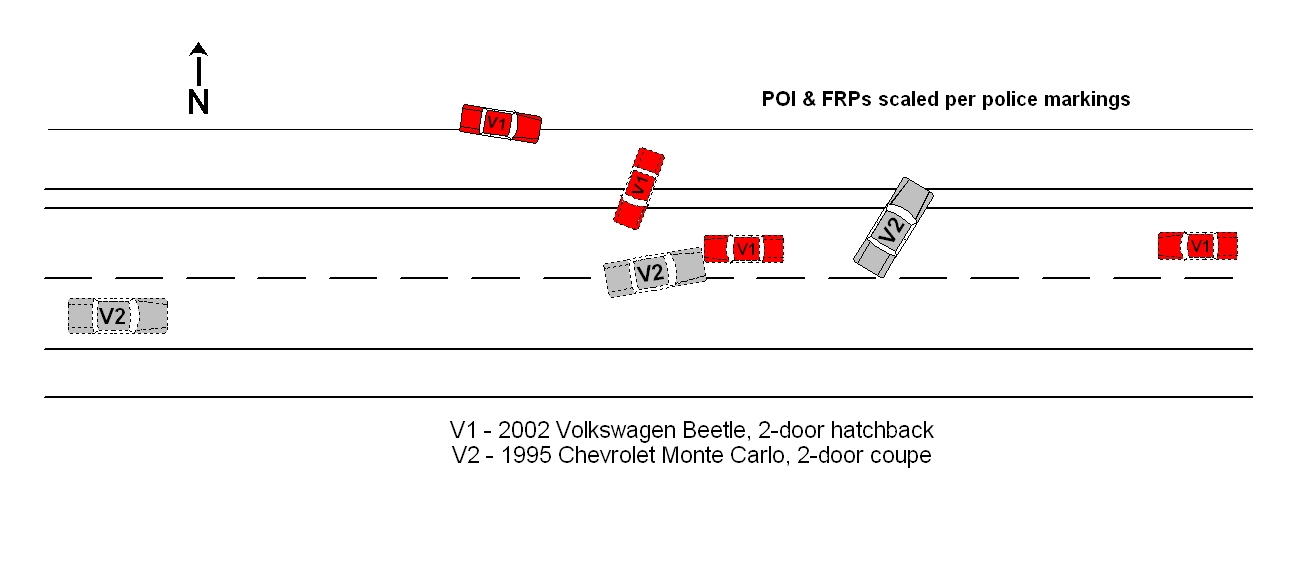}
    \label{fig:128697_sketch}}
    \hspace{1cm}
    \subfloat[Original summary of scenario 128697.]{\includegraphics[width=10cm]{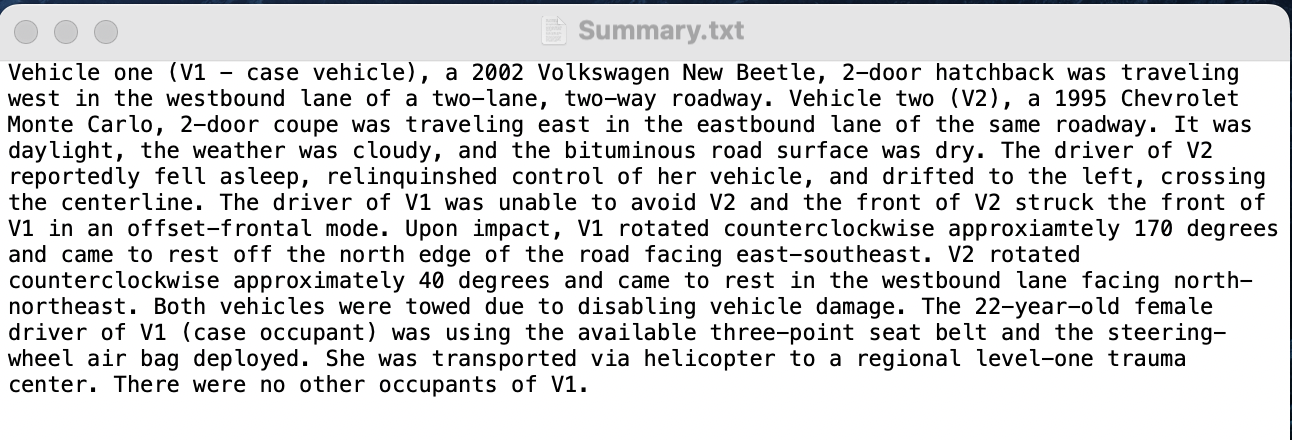}
    \label{fig:128697_summary}}
    \captionsetup{font=small}
    \caption{Visual comparison of simulated scene, original crash sketch, and summary for scenario 128697.}
    \label{fig:fidelity_check}
\end{figure}

To answer RQ2, the validated Extended Scenic DSL was converted into executable Scenic scripts and used to reconstruct crash scenarios in CARLA 0.9.15. The reconstructed scenarios were then evaluated in the Autoware based simulation setup through both the CARLA Autoware ROS bridge and the Zenoh bridge. An initial visual fidelity check was also performed, comparing simulated scenes against original police sketches and narrative summaries (Figure~\ref{fig:fidelity_check}). The comparison confirmed alignment in road layout, lane structure, weather, time, vehicle placement, and collision configuration.

\begin{figure}[!ht]
    \centering
    \subfloat[Straight-1 sample.]{\includegraphics[width=6cm]{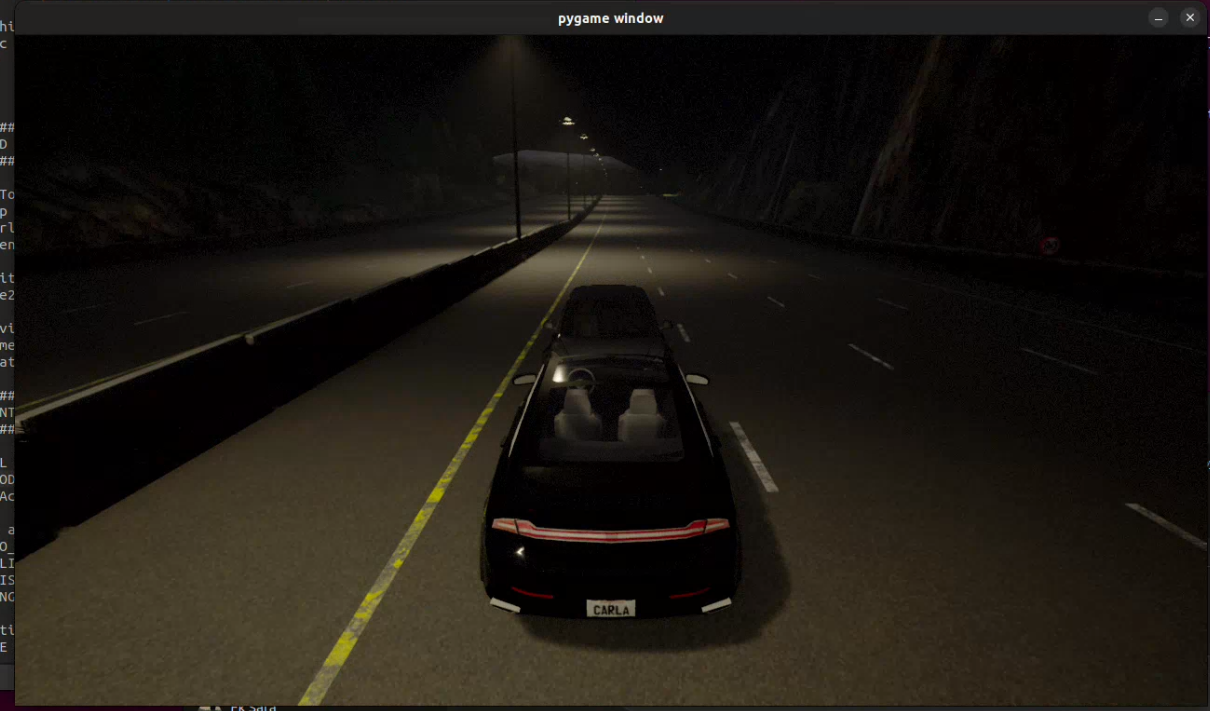}
    \label{fig:straight_2_collsion}}
    \hspace{1cm}
    \subfloat[Straight-2 sample.]{\includegraphics[width=6cm]{./figs/carla_straight.png}
    \label{fig:carla_straight}}
    \hspace{1cm}
    \subfloat[Intersection-1 sample.]{\includegraphics[width=6cm]{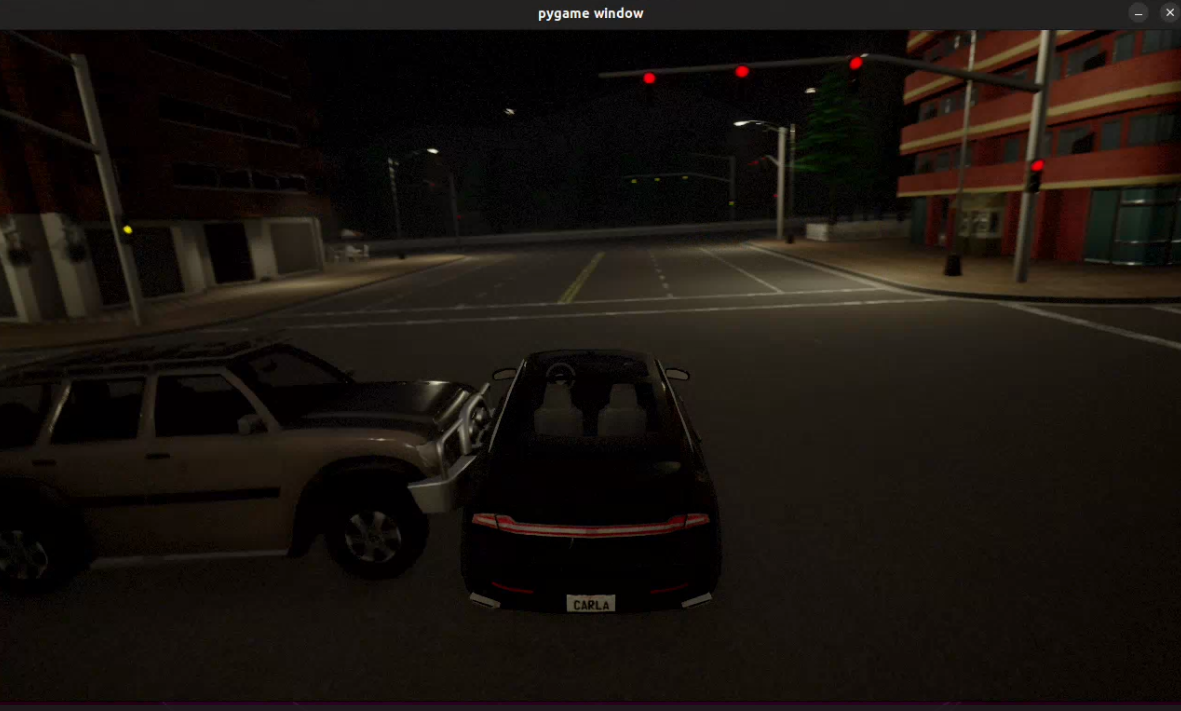}
    \label{fig:intersection_colliosion}}
    \hspace{1cm}
    \subfloat[Intersection-2 sample.]{\includegraphics[width=6cm]{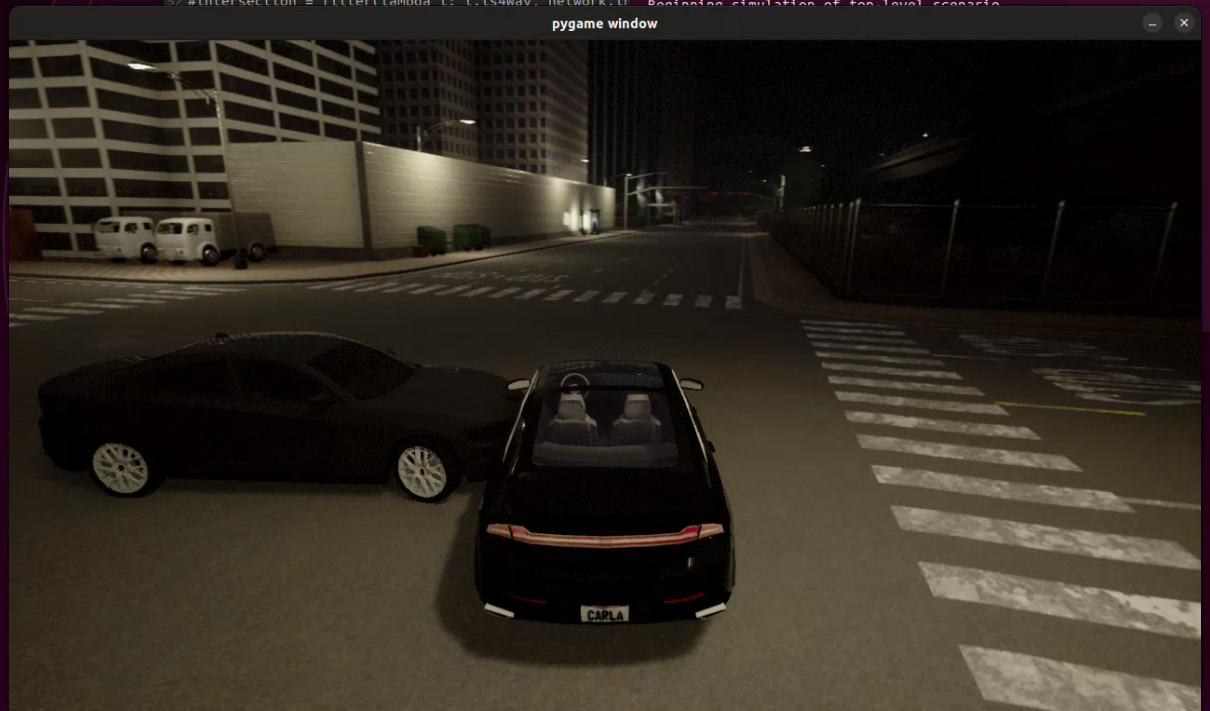}
    \label{fig:intersection_2_collsion}}
    \hspace{1cm}
    \subfloat[T-intersection sample.]{\includegraphics[width=6cm]{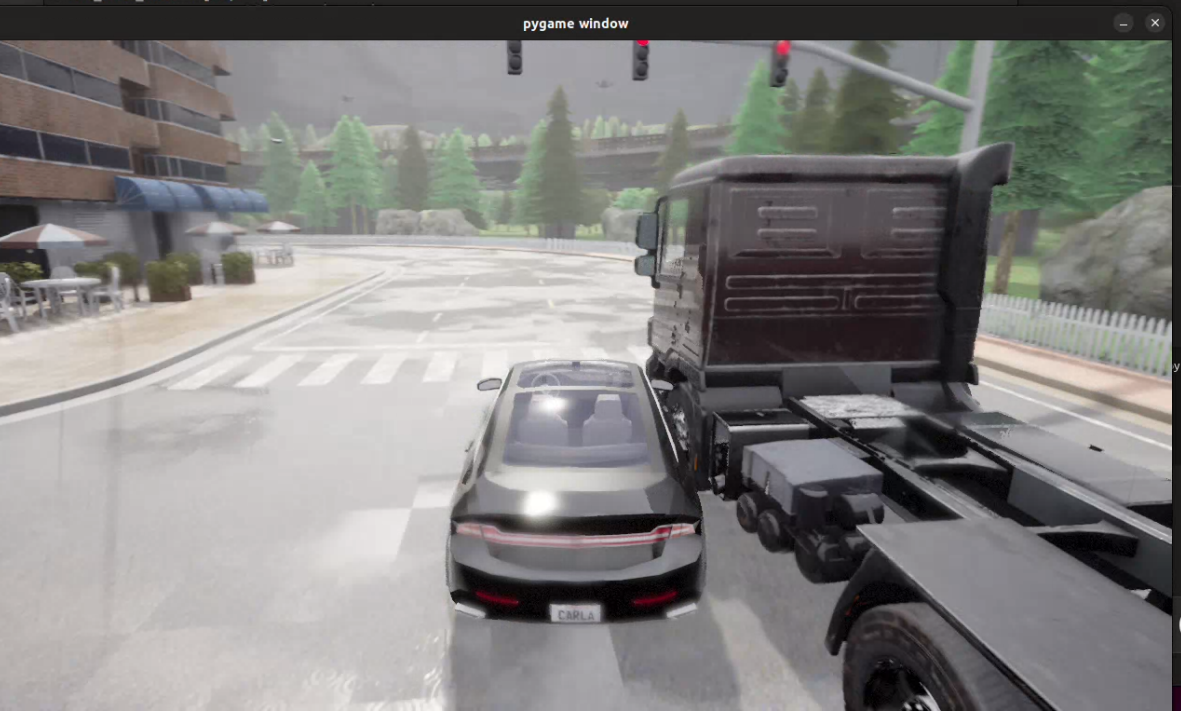}
    \label{fig:T-intersection_collsion}}
    \hspace{1cm}
    \subfloat[Curve sample.]{\includegraphics[width=6cm]{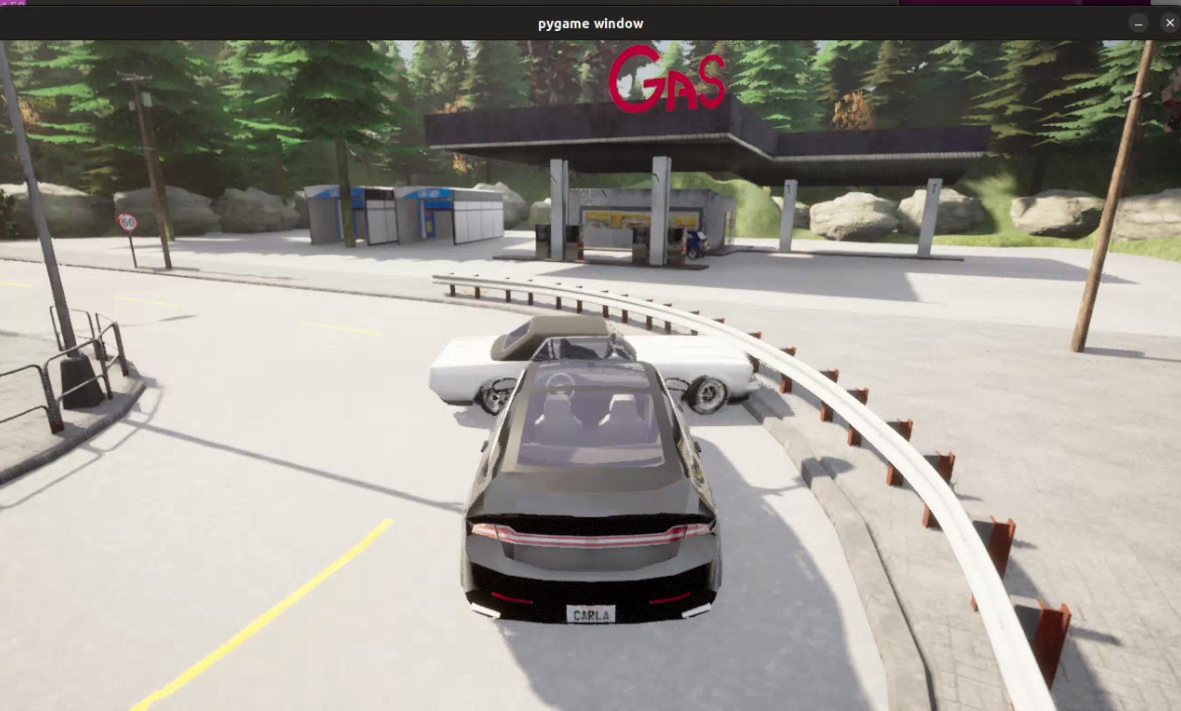}
    \label{fig:curve_collision}}
    \captionsetup{font=small}
    \caption{Examples of reconstructed scenarios examined by human assessor.}
    \label{fig:human_check}
\end{figure}

For robustness, the study used Scenic probabilistic operators to generate 2,000 randomized instantiations per scenario while preserving the core crash geometry. From these, 20 representative samples were selected for human evaluation, with examples shown in Figure~\ref{fig:human_check}. Two independent assessors reviewed the reconstructed scenarios and rated fidelity using a four point scale consisting of \textit{Match}, \textit{Mostly Match}, \textit{Partially Match}, and \textit{No Match}. This evaluation design is important because fidelity is a human judgement task, and independent assessment is needed to reduce bias and support an agreement based interpretation of the results.

\begin{figure*}[!ht]
	\setlength{\belowcaptionskip}{-0.3cm}
	\centering
	\includegraphics[width=0.9\textwidth]{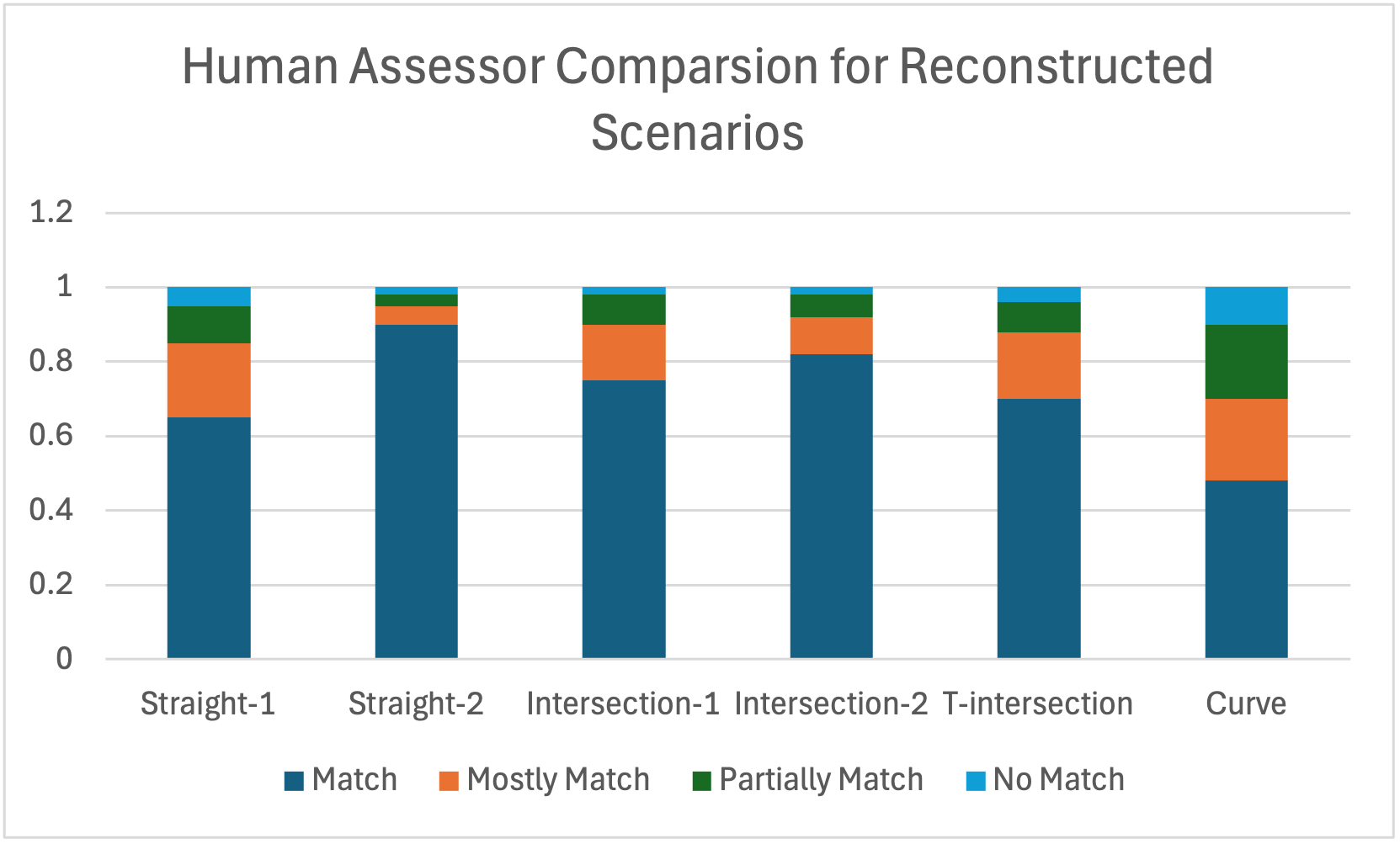}
	\caption{Human assessor comparison for reconstructed scenarios.}
	\label{fidelity}
\end{figure*}

The fidelity results (Figure~\ref{fidelity}) show that the reconstructed scenarios preserve the semantics of the original crash reports well, especially for structured road topologies. Straight-2 and Intersection-2 scenarios achieved the highest fidelity, with more than 80\% of ratings in the \textit{Match} category. T intersection scenarios also performed strongly, with the combined \textit{Match} and \textit{Mostly Match} categories accounting for nearly 90\% of responses. Overall, these results show that the proposed pipeline can reconstruct semantically faithful scenarios, while also preserving enough structure for systematic testing.

\subsection{RQ3: Testing Utility}
To answer RQ3, we executed the generated Scenic scripts in the Autoware CARLA simulation environment using representative scenarios across the main road topologies identified during data collection. The selected scenarios cover Straight, Intersection, T intersection, and Curve configurations, which ensures that the evaluation spans different geometric constraints and traffic rule dependencies. The study also used Scenic probabilistic programming to vary key parameters such as starting positions, velocities, and environmental conditions while keeping the semantic constraints of the original crash descriptions unchanged.

The utility of the generated scenarios is supported by the agreement between automated failure detection and human assessment. We compared Autoware detected rule violations against manual counts from human assessors and reports strong alignment across topologies. In most cases, the automated pipeline reproduced the same number of violations expected by human assessors. Table~\ref{tab:rule_violation} shows exact agreement for Straight 1, Intersection 2, T intersection, and Curve, and near agreement for Straight 2 and Intersection 1, where one assessor reported one fewer violation than the oracle count. This result is important because it shows that the generated scenarios are not only geometrically valid, but also functionally effective for safety testing.

\begin{table}[!ht]\scriptsize
\centering
\caption{Comparison of Human Assessors and Oracle Rule Violation Counts}
\begin{tabular}{|l|c|c|c|}
\hline
\textbf{Road-Type} & \textbf{Human Assessor 1} & \textbf{Human Assessor 2} & \textbf{Oracle Rule Violation Count} \\
\hline
Straight-1        & 1 & 1 & 1 \\
Straight-2        & 2 & 1 & 2 \\
Intersection-1    & 3 & 2 & 3 \\
Intersection-2    & 3 & 3 & 3 \\
T-intersection    & 2 & 2 & 2 \\
Curve             & 2 & 2 & 2 \\
\hline
\end{tabular}
\label{tab:rule_violation}
\end{table}

The scenario level examples further show the testing value of the approach. In the straight‑road scenarios, the pipeline reproduced unsafe overtaking and wrong‑way driving cases, including a centerline crossing and a head‑on collision context under low visibility, with rule violations related to lane discipline, overtaking safety, and low‑visibility requirements. In the intersection scenarios, the generated scenes induced failures related to red light compliance, stop sign obligations, and right of way, while preserving the collision geometry described in the original reports. In the T intersection case, the reconstructed scenario correctly prioritized the crash narrative of brake failure over the normal expectation of stopping, which forced entry into the intersection and reproduced the documented T bone collision pattern. In the curve scenario, the generated scene preserved both road curvature and actor deviation, and it reproduced a head-on collision at the curve apex together with lane discipline and red-light violations. These examples show that the proposed pipeline supports realistic and safety relevant testing across diverse traffic situations.

\section{Related Work}\label{sec:related_work}
Research on ADS testing has developed along two main directions: scenario-based testing and search-based testing. Scenario-based testing focuses on building representative driving situations and checking whether the ADS satisfies safety requirements in those situations. Frameworks such as Matrix-Fuzzer \cite{zhang2023realistic} and ISS-Scenario \cite{li2024iss} show how this idea can be implemented in practice. Matrix-Fuzzer combines behavior abstraction and optimization to generate safety critical scenarios, while ISS-Scenario supports configurable batch testing and replay in simulation. In contrast, search-based testing treats test generation as an optimization problem and uses fitness functions to find challenging inputs. Frameworks such as OpenSBT \cite{sorokin2024opensbt} and RIGAA \cite{humeniuk2024reinforcement} illustrate this line of work by combining simulators with optimization methods, including evolutionary search and reinforcement learning-guided search. These studies provide strong foundations for ADS testing, but they usually assume that scenarios or parameter spaces are already available in a structured form.

A second line of work focuses on scenario representation and domain specific languages for simulation-based testing. Existing DSLs such as OpenSCENARIO \cite{OpenSCENARIO}, GeoScenario \cite{queiroz2019geoscenario}, SML4ADS \cite{li2022sml4ads}, and SceML \cite{schutt2020sceml} support structured scenario specification, but they differ in scope, usability, and simulator dependence. Some of them are strong in map and route modeling, while others are designed for visual modeling or planning oriented tests. Scenic \cite{fremont2019scenic} is particularly relevant to this work because it provides a probabilistic programming language for scenario generation, supports geometric constraints, and can represent uncertainty through sampling. This makes Scenic suitable for generating diverse scenario instances from a compact specification. However, writing Scenic programs still requires domain knowledge and careful control of behavior definitions, which creates a barrier for direct use with unstructured crash reports.

Recent studies have started to generate ADS test scenarios directly from crash reports and other real world sources. AC3R \cite{gambi2019generating} is an early framework that reconstructs crash scenarios from police reports using NLP and an ontology-based pipeline, then converts the extracted information into simulation-based test cases. It showed the feasibility of report-driven scenario generation, but it relied on text processing pipelines and had difficulty with long reports and scene realism. CRISCE \cite{nguyen2022crisce} moved to sketch driven reconstruction and used image processing to extract road layout, vehicle locations, and trajectories from accident sketches. This improved the use of visual crash evidence, but its performance depends on image quality and manual pre-processing. LCTGen \cite{tan2023language} introduced LLM-based text conditioned traffic generation with a pipeline that includes an interpreter, a generator, and a retrieval module. It improved controllability and realism for text-based crash descriptions, but it did not use crash sketches as a core input and did not perform explicit validation of the generated scenario representations.

ScenicNL \cite{elmaaroufi2024scenicnl} further advanced this area by converting natural language crash reports into probabilistic Scenic programs. It combined prompting strategies and constrained generation to improve syntax correctness and produce executable scenarios with uncertainty. This is an important step toward probabilistic scenario synthesis from crash narratives. However, a semantic consistency issue remains between the generated programs and the original crash descriptions. TARGET \cite{Deng2023TARGETTR} addresses a different but related problem by deriving test scenarios from formal traffic rules. It uses an LLM to parse rules into a compositional DSL, validates syntax and semantics at the component level, and then generates executable scripts through map-based search and parameter mapping. TARGET provides strong modularity and reliable rule-grounded testing, but it is designed for deterministic rule-based scenario reproduction and does not directly model the stochastic variability that is common in real crash reports.

Overall, existing work provides key building blocks for automated ADS testing, including scenario generation frameworks, DSL-based representations, and multimodal crash reconstruction methods. However, a gap remains between multimodal crash report understanding and probabilistic, executable scenario synthesis that also supports systematic validation. This gap motivates our use of an Extended Scenic DSL and an LLM-based multimodal knowledge extraction and validation pipeline, which aims to preserve semantic fidelity while enabling scalable scenario variation for ADS testing.

\section{Discussion}\label{sec:discussion}
This study extends the TARGET line of work from deterministic scenario reconstruction to probabilistic scenario testing. The key advance lies not in better script generation, but in a shift in testing logic. The Extended Scenic DSL acts as a semantic and syntactic constraint layer, and the Scenic-based pipeline supports probabilistic variation and repeated search driven testing rather than a single replay of one crash report. This design supports stable extraction accuracy, strong scenario fidelity, and practical failure discovery in the ADS testbed.

A key element is the choice of Autoware and the test infrastructure built around it. This work highlights the bridge system as a testing infrastructure milestone, and explains that the combination of the CARLA-Autoware ROS bridge with the Zenoh bridge makes the Scenic generation pipeline operational for large scale ADS testing. This integration reduces setup friction and enables batch execution of many generated Scenic scripts in a more practical workflow. We also presented the Zenoh bridge setup as part of the experimental testbed configuration, which supports reliable communication between the simulation side and the Autoware side. This strongly motivated the selection of Autoware in this work, particularly given the goal of scalable, scenario‑based testing for ROS‑based software.

To clarify the ecosystem positioning, Autoware was selected because only a few mature autonomous driving stacks are practical for this level of end‑to‑end testing research. Autoware and Baidu Apollo \cite{Apollo} are the most prominent examples in the literature and prior benchmarking practice \cite{deng2022scenario}. This positioning aligns with TARGET, which evaluated both Autoware and Apollo in its ADS experiments. It also explains why this work prioritizes a robust Autoware‑based integration and testing pipeline over broad platform coverage in an initial implementation.

A second discussion theme is model choice. Unlike TARGET, which explicitly compares several LLMs for rule parsing, this study uses a single model, GPT‑4o mini, and still achieves comparable practical outcomes through a stronger pipeline design, constrained DSL representation, and a validation stage. This is a key engineering insight: it shifts the focus from model size to system design quality, particularly in token budgeting and cost efficiency. In other words, this work demonstrates that a compact model suffices when the representation schema and validation process are well designed.

A third discussion theme is generalization. This work does not rely on handcrafted edge cases. Instead, it uses random samples from original crash derived scenarios together with a representative set of traffic rules, then evaluates them across road topology classes. This sampling strategy tests whether the pipeline generalizes beyond a few curated examples. The selected cases span diverse topological road structures, reinforcing that the fidelity and utility findings are not limited to a single map pattern. The probabilistic Scenic representation further improves this generalization because each script can generate many valid variants under the same semantic constraints.

For this paper, inter-rater agreement should be emphasized as a validity anchor for both representation accuracy and scenario fidelity. The experiment reported high agreement among evaluators, which supports the use of human ratings as a credible reference for semantic correctness and scenario faithfulness. This point aligns well with TARGET, which explicitly explains the interpretation of Fleiss kappa and uses agreement thresholds to justify human evaluation quality. By making this link explicit, the paper strengthens the methodological claim that the human‑derived oracle is reliable, not merely convenient.

Finally, the discussion highlights the testing utility in practical terms. The probabilistic test generation not only reproduced expected violations but also exposed additional failure behaviors in Autoware that were not explicitly anticipated by human reviewers, including involuntary violations caused by control‑level failure conditions. This is the main benefit of moving from deterministic reconstruction to probabilistic search-based testing. It shows that the pipeline can preserve scenario semantics while still expanding the failure surface of the ADS under realistic variation.

\section{Conclusion}\label{sec:Conclusion}
This paper presents an LLM‑based pipeline that generates executable autonomous driving system test scenarios from multimodal crash reports. The method combines GPT‑4o mini, a validated knowledge parsing process, and an Extended Scenic DSL to convert crash summaries and sketches into structured scenario representations and subsequently into Scenic programs for simulation. Unlike deterministic rule‑reproduction pipelines, the proposed approach supports probabilistic scenario generation, enabling systematic variation of key scene elements and broader testing coverage under consistent crash semantics. Experiments demonstrate high representation accuracy against a human‑derived golden oracle, strong scenario fidelity in reconstructed simulations, and clear testing utility in an Autoware and CARLA environment, where generated scenarios reproduce traffic rule violations and expose safety‑relevant failures. These results suggest that the proposed pipeline offers a practical and scalable method for linking real crash evidence with simulation‑based ADS testing. Future work may extend the rule set and crash coverage, improve parameter search for more efficient failure discovery, and evaluate the pipeline on additional industry‑grade platforms such as Apollo.



 \bibliographystyle{elsarticle-num} 
 \bibliography{cas-refs}





\end{document}